\documentclass[showpacs,prb,aps,twocolumn]{revtex4}
\usepackage{graphicx}
\begin{document}
%
\title{Protein domain connectivity and essentiality }
\author{
L.\ da\ F.\ Costa, F.\ A.\ Rodrigues\ and G.\ Travieso }
\affiliation{ Instituto de F\'{\i}sica de S\~{a}o Carlos,
Universidade de S\~{a}o Paulo, PO Box 369, 13560-970, S\~{a}o
Carlos, SP, Brazil }
\begin{abstract}

Protein-protein interactions can be properly modeled as scale-free
complex networks, while the lethality of proteins has been correlated
with the node degrees, therefore defining a lethality-centrality
rule. In this work we revisit this relevant problem by focusing
attention not on proteins as a whole, but on their functional domains,
which are ultimately responsible for their binding potential. Four
networks are considered: the original protein-protein interaction
network, its randomized version, and two domain networks assuming
different lethality hypotheses. By using formal statistical analysis,
we show that the correlation between connectivity and essentiality is
higher for domains than for proteins.
\end{abstract}


\maketitle

A great deal of the functionality of proteins stems from their
ability to dock, i.e. to connect.  Such dockings are highly specific
and depend on geometrical and field compatibilities between the
involved proteins. More specifically, the docking sites of a protein
are largely defined by the presence of specific \emph{domains}, i.e.
portions of aminoacid sequences along the protein primary
backbone~\cite{Ng03}. Given that protein-protein interactions
involve physical interactions between protein domains, domain-domain
interaction information can be particularly useful for validating,
annotating, and even predicting protein interactions. The subject of
protein domain interaction has been covered in previous
investigations~\cite{Wuchty01,Deng02:Genome}.

Protein-protein interaction networks are obtained by representing
each protein as a node and each possible docking between pairs of
proteins as edges linking the respective nodes. Domain-domain
interaction networks are constructed considering protein complexes,
Rosetta Stone sequences, and by using protein interaction
networks~\cite{Wuchty01,Wuchty02,Ng03}. The current work considers
the last approach, taking into account domain subnetworks contained
in protein-protein interaction networks.  This method allows not
only the direct visualization of the coexistence of domains and
proteins, naturally providing for the multiplicity of domains, but
also the objective quantification of interactions between domains.

Given a network, the degree of a specific node is defined as the
number of connections between that node and the remainder of the
network. This frequently used measurement can be generalized to
express the connectivity not of a single node, but of a whole
subnetwork contained in the original network~\cite{Costa:2006}.
Subnetworks can be obtained by selecting a subset of nodes from the
original network as well as the edges between those nodes. The
degree of a subnetwork is then defined as the number of connections
between its nodes and the remainder of the network nodes, not taking
into account the connections internal to the subnetwork.  By
quantifying the number of interactions between the subnetwork and
the overall structure, the subnetwork degree provides a valuable
indication about the role and importance of each subnetwork.

\begin{figure}[h]
\centerline{\includegraphics[width=\columnwidth]{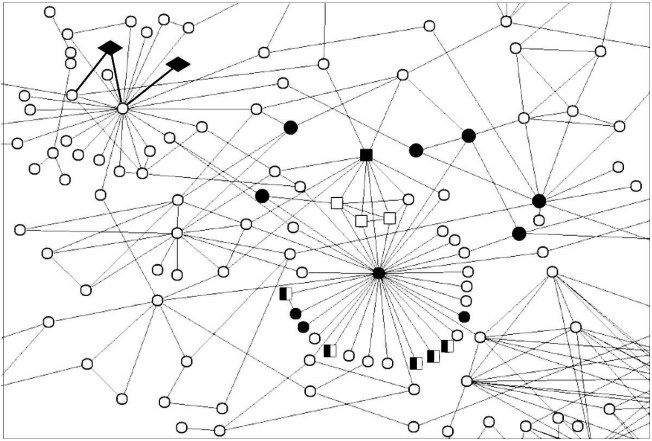}}
\caption{Partial representation of the \emph{S. cerevisiae}
protein-protein interaction network, with four domain subnetworks
identified in black circles, white squares, black squares and black
diamonds. The degree of the latter can be obtained by counting the
number of edges between the black diamond nodes and the other nodes
of the structure, yielding total degree equal to $3$. Note that the
presence of overlap between the white squares subnetworks and black
squares subnetworks, characterized by the five black/white squares
indicates the presence of two different domains in the proteins
associated to those nodes.}\label{fig1}
\end{figure}

One particularly interesting way to define a subnetwork is by
selecting among the nodes in the original network those that exhibit
some specific feature.  Considering a protein-protein interaction
network, a subnetwork can be obtained by selecting those nodes that
contain one or more instances of a specific protein domain. Note
that such a subnetwork is embedded within the original
protein-protein interaction network. A whole collection of
subnetworks can then be obtained, one for each considered domain,
and valuable insights about the importance and role of the domains
can be inferred by using the concepts of subnetwork degree and
subnetwork hubs.  We applied such concepts to the
\emph{Saccharomyces cerevisiae} protein-protein interaction networks
using the non-redundant database of interacting proteins by Sprinzak
\emph{et al.}~\cite{Sprinzak03} (formed by $4,135$ proteins and
$8,695$ connections). The respective domains were identified by
using the Pfam database~\cite{pfam}, which contains a large
collection of multiple sequence alignments and profile hidden Markov
models (HMM) covering the majority of protein domains, yielding a
total of $1,424$ domains. Figure~\ref{fig1} shows part of such a
network, where four domain subnetworks are identified in black
circles, white squares, black squares and black diamonds.

In order to investigate the relationship between domain connectivity
and lethality, it is necessary to extend the concept of essentiality
to domains. However, as there is no consensus about domain lethality
in the scientific literature, we suggest the two following
hypotheses:

\begin{itemize}
  \item I. \emph{Domain lethality in a weak sense}: a domain is lethal if
  it appears in a lethal protein.
  \item II. \emph{Domain lethality in a strong sense}: a domain is lethal
  if it appears in a single-domain lethal protein.
\end{itemize}

\begin{table}[h]
\caption{Statistical values for protein and domain networks.}
\begin{tabular}{c|c|c|c|c|c|c|c}
  & $N$ &$N_{L}$ & $\langle k \rangle$ & $k_c$ & $\gamma$ & $r$ & $\rho$ \\
  \hline
    Proteins                & 4,135 & 795 & 2.10 & 38 & 2.8 & 0.12 & 0.10 \\ \hline
    Domains in weak sense   & 1,424 & 499 & 2.24 & 22 & 2.5 & 0.61 & 0.61 \\ \hline
    Domains in strong sense & 818   & 243 & 1.27 & 15 & 2.6 & 0.73 & 0.77 \\ 
\end{tabular}\label{table1}
\end{table}

\begin{figure}
\centerline{\includegraphics[width=\columnwidth]{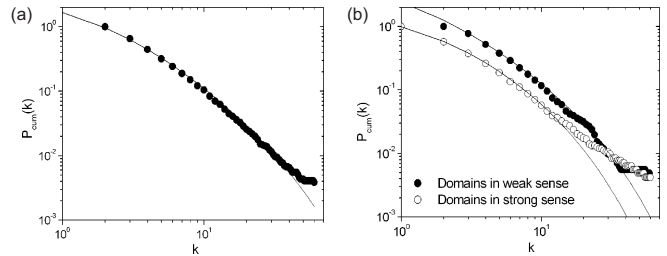}}
\caption{The cumulative degree distribution of proteins and domain
subnetworks in the considered \emph{S.~cerevisiae} follow a
power-law with an ex\-po\-nen\-ti\-al cut-off,
$\mathrm{P(k)\approx(k + k_0)^\gamma e^{-(k+k_0)/k_c}}$, represented
by the continuous line. The values of $k_c$ and $\gamma$ for
proteins and domains are presented in
Table~\ref{table1}.}\label{fig2}
\end{figure}

The first definition is considered weak because a lethal domain can
appear in lethal and viable proteins simultaneously. However, that
assumption is still potentially interesting because co-occurring
domains are more likely to exhibit similar function or localization
than domain in separate proteins~\cite{Ng03,Vogel05}, which suggests
that lethal proteins may involve uniformly lethal domains. The second
hypothesis, on the other hand, is considered strong because if the
domain is the only one in a lethal protein, it must be responsible for
the protein's essential function. When working with the first
assumption, the whole protein interaction network is studied; for the
second assumption, only the subnetwork formed by proteins with a
single domain is considered. It is important to note that the two
lethality situations above are just hypotheses to be checked against
the experimental results reported by Jeong \emph{et
al.}~\cite{Jeong:2001} concerning protein-protein interaction
networks. In other words, eventual identification of high correlation
between degree and lethality for one of those hypotheses could be
understood as supporting that respective assumption due to the
\emph{centrality-lethality rule}, which is widely believed to reflect
the special importance of hubs in organizing the network, and the
biological significance of network architectures, a key notion in
systems biology~\cite{He06:PLOS}. Figure~\ref{fig2} shows the
histogram of the cumulative protein degree and domain subnetworks
degrees in both weak and strong senses. The cumulative degree
distribution for all networks follows a power law with an exponential
cut-off~(finite size effect~\cite{Newman05CP}) described by
$\mathrm{P(k)\approx(k + k_0)^\gamma
e^{-(k+k_0)/k_c}}$~\cite{Jeong:2001}. The values of $k_c$ and $\gamma$
obtained from the cumulative distribution for proteins and domains are
presented in Table~\ref{table1}.

\begin{figure}[h]
\begin{center}
\centerline{\includegraphics[width=\columnwidth]{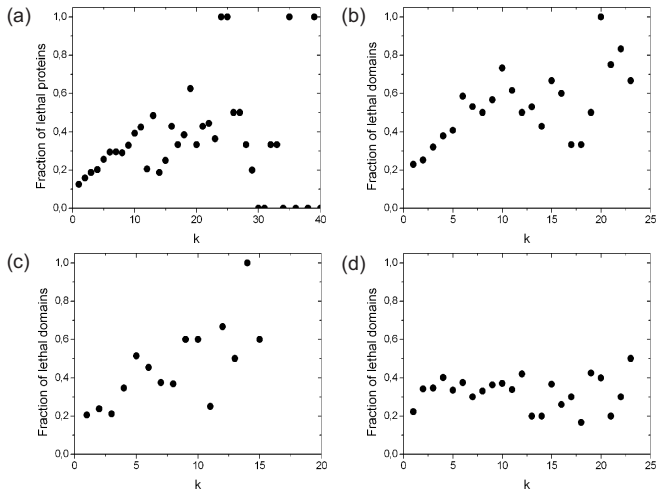}}
\caption{Fraction of lethal proteins and domains with a particular
degree for (a) proteins, (b) domains in the weak sense, (c) domains
in the strong sense and (d) the network using protein permutations
(see the text for details).}\label{fig3}
\end{center}
\end{figure}

Figure~\ref{fig3} shows the relationship between degree and
essentiality for the protein and domain networks. The lethality of
proteins was determined using the MIPS database~\cite{mips} and the
number of lethal protein, $N_L$, for the considered networks is shown
in Table~\ref{table1}. The abscissae represents the node degree $k$ of
proteins or domains\footnote{The domain degree is normalized by the
number of proteins present in the subgraph of the respective domain so
as to avoid artificially high degree otherwise induced by more
abundant domains, which would bias the results.} limited by the
cut-off (see Figure~\ref{fig2}) whose values are presented in
Table~\ref{table1}, while the ordinate axis expresses the fraction of
lethal proteins or domains among the ones with degree $k$. In order to
determine the correlation between the fraction of lethal
proteins/domains and their degree, we estimated the Pearson
correlation coefficient~\cite{Edwards1976},$r$, which measures the
strength of linear relationship between two variables, and the
Spearman rank correlation coefficient~\cite{Lehmann98,Freund93},
$\rho$, which is a nonparametric coefficient used in case of nonlinear
relationships. Table~\ref{table1} presents the values of correlations
for proteins and domains, which indicates that the correlation between
lethality and degree is larger for hypotheses (I) and (II) than for
whole proteins. The statistical significance of the correlations was
tested by applying the Fisher's comparison correlation coefficient
test~\cite{Fisher21}, $p$. The comparison between the correlation
coefficients of proteins and domains in weak sense results $p \leq
0.035 $ for $r$ and $p \leq 0.001$ for $\rho$.  The comparison of
proteins with domains in weak sense yields $p \leq 0.015$ for $r$ and
$p \leq 0.001$ for $\rho$. These results lead to the conclusion that
the domains correlations in both weak and strong sense are
significantly higher than the correlation obtained for proteins.

Since protein domains represent the basic evolutionary units that form
proteins, it is not surprising that domains should play a fundamental
role in the definition of proteins interaction and
lethality~\cite{Vogel05}. In this way, the obtained results indicate
that the interactions between proteins may be defined at the domain
level, with the importance of domains being associated to their
functions~\cite{Ng03}. As hubs tend to be the most important nodes in
networks, domains with larger number of connections should be
particularly fundamental (essential) for network maintenance.  Indeed,
domains with a high number of connections act as interconnecting
pathways in the network (also called backbones) which, when removed,
imply substantial network diameter increasal (as discussed by
Jeong~\emph{et al.}~\cite{Jeong:2001}).  The special importance of
domain essentiality can be readily inferred by inspecting the results
presented in Figure~\ref{fig3} and Table~\ref{table1}. Further, lethal
domains are more likely to be hubs than lethal proteins. In other
words, both hypotheses about domain lethality have been supported by
the experimental results, with the strong sense hypothesis resulting
more definite than the weak sense counterpart.

In order to verify whether the distribution of domains among
proteins influences the domains connectivity, we randomized the
protein positions along the network while maintaining the network
structure, which was done by permutations of the proteins assigned
to the nodes. Thus, for $100$ randomized network versions (see in
Figure~\ref{fig3}(d)), the correlations obtained are close to zero
(see Table~\ref{table1}), confirming that the relation between the
connectivity and lethality is unlikely to be a spurious effect.

The results presented here suggest a novel fundamental relationship
between protein and domain interaction which has several
implications for future works, as validating, annotating, and even
predicting protein interactions and lethality. Also, our results can
be used as a pre-investigation to obtain experimental data about
domain interaction and lethality.

The authors thank E. Sprinzak for the protein interaction data.
Luciano da F. Costa is grateful to FAPESP (proc. 99/12765-2), CNPq
(proc. 308231/03-1) and the Human Frontier Science Program
(RGP39/2002) for financial support. Francisco A. Rodrigues
acknowledges FAPESP sponsorship (proc. 04/00492-1).


\end{document}